# A fast field-cycling MRI relaxometer for physical contrasts design and pre-clinical studies in small animals

**Javier A. Romero, Gonzalo G. Rodriguez and Esteban Anoardo***

*Laboratorio de Relaxometría y Técnicas Especiales (LaRTE), Grupo de Resonancia Magnética Nuclear. FaMAF - Universidad Nacional de Córdoba e IFEG-CONICET, Córdoba – Argentina*

*Corresponding Author:* Esteban Anoardo. Famaf - UNC/IFEG - CONICET. Ciudad Universitaria. Córdoba, Argentina. **anoardo@famaf.unc.edu.ar**

**Abstract**

We present a fast field-cycling NMR relaxometer with added magnetic resonance imaging capabilities. The instrument operates at a maximum proton Larmor frequency of 5 MHz for a sample volume of 35 mL. The magnetic field homogeneity across the sample is 1400 ppm. The main field is generated with a notch-coil electromagnet of own design, fed with a current whose stability is 220 ppm. We show that images of reasonable quality can still be produced under such conditions. The machine is being designed for concept testing of the involved instrumentation and specific contrast agents aimed for field-cycling magnetic resonance imaging applications. The general performance of the prototype was tested through localized relaxometry experiments, $T_1$-dispersion weighted images, temperature maps and $T_1$-weighted images at different magnetic field intensities. We introduce the concept of positive and negative contrast depending on the use of pre-polarized or non-polarized sequences. The system is being improved for pre-clinical studies in small animals.

1. Introduction

During the last years, the development of Magnetic Resonance Imaging (MRI) scanners operating at magnetic fields greater than 3T was highly motivated by special applications like neuroscience [1] and fast imaging [2], among others. Such magnetic field strengths are commonly reached by using superconducting magnets working at fixed Larmor frequencies. In the case of $T_1$ (spin-lattice relaxation time) -weighted contrast, fixed-field MRI may present a limitation in acquiring images with maximal contrast: maximum $T_1$ differences occur at specific frequencies depending on sample composition [3]. This has previously been noticed for $T_1$-weighted contrast for human white and gray matter [4], for example, where maximal contrast is given at 10 MHz (proton Larmor frequency). On the other hand, working at fixed, low-magnetic field has a negative impact on the signal-to-noise ratio (SNR). Therefore, it can be said that although fixed-field MRI is a robust and highly-developed technique, the contrast based exclusively on $T_1$ relaxation times may be suboptimal. The use of high magnetic fields and other physical contrast mechanisms like other relaxation times, diffusion, susceptibility or chemical contrast agents partially compensate the mentioned limitation, but also enhance other physical problems like artifacts associated with subject motion [5,6], the electromagnetic wavelength of the associated radio-frequency (RF) fields used for nuclear spin manipulation [7], more complex risk management [8] and increasing costs [9,10], among other technical limitations that are still a matter of engineering refinement.

A potential solution to the suboptimal $T_1$-weighted imaging at high fixed-fields relies on the possibility to cycle the magnetic field between different values. This practice allows the spin-system to evolve at lower magnetic field intensities, where the difference in $T_1$ between heterogeneous components of the sample usually increases. This feature opens new possibilities for physical contrast mechanisms that are impermissible to fixed high-field MRI. Cycling of the magnetic field is a well-known and deeply studied technique for measuring the $T_1$ dependence on the Larmor frequency ($T_1$-dispersion curves), as well as for double-resonance spectroscopy involving quadrupolar nuclei or electrons (also known as fast field-cycling or FFC) [11-14]. Field-cycling nuclear magnetic resonance (NMR) experiments may demand rapid variations between different magnetic field intensities, so electromagnets are usually involved. Technological developments on this NMR sub-field are oriented towards working at lower magnetic fields, with higher requirements on the magnetic field homogeneity and stability, improving magnet switching performance and boosting the SNR [15]. Experiments may be achieved by polarizing the spin system in a high intensity magnetic field, then quickly switching down to a low intensity field for a pre-established evolution period, and finally bringing the field intensity back to a higher value for acquisition (magnetic field cycling sequence referred to as "pre-polarized" or PP). Alternatively, the spin-system may be allowed to evolve after switching-up the magnetic field, starting from a non-polarized state (non-polarized or NP) [13].

The combination of FFC and MRI, namely FFC-MRI, is a technique which allows acquisition of images of a spin system that has evolved at different magnetic field intensities. Probably firstly suggested by Yamamoto et al. [16], cycling an auxiliary magnetic field was successfully used to boost the SNR in MRI experiments. This additional magnetic field was applied only during a pre-polarization period in order to increase the sample magnetization prior to the application of the MRI pulse sequence. Cycling the magnetic field was successfully used for proton-electron double resonance imaging (PEDRI) [17]. A specialized system based on this technique did not appear until a few years later [18]. Groups from the USA and Canada have also contributed with technological developments in this area [19-22]. All these FFC-MRI examples (and subsequent work, see for example [23-27]) were mainly thought from an MRI perspective. Even the concept of relaxometric imaging was originally approached from a pure MRI point of view, using a whole-body magnetic resonance imager [28,29]. In almost all the cases an insert or combined polarizing magnet was switched in combination with a static readout homogeneous field. We present here the alternative approach of employing a single electromagnet driven by a field-cycling relaxometer, upgraded with all the needed hardware for image acquisition: the MRI-relaxometer.

The possibility to image samples that may evolve at a wide range of magnetic field strengths offers advantages unreachable by conventional MRI scanners with fixed $B_0$ field.

One example is the acquisition of protein images through use of quadrupolar dips [22,30]. However, other possibilities can be explored within this scheme to produce images weighted by different physical parameters. For instance, spin-system manipulations at ultra low field (or any non-zero magnetic field within the range allowed by the available hardware) are still unexplored options.

In this work, the performance of our prototype device was analyzed through localized relaxometry and $T_1$ dispersion-weighted imaging experiments. In addition, we show that even using a low-homogeneity magnetic field, the system can be used for the study and design of contrast mechanisms. This last point has a direct impact in the cost of the associated hardware, thus providing an attractive solution for research and development in the field. The studied contrast mechanisms and developed protocols can be later tested in a whole-body field-cycling MRI scanner for specific diagnosis.

A word of caution should be mentioned concerning MRI at moderate fields and homogeneities. The challenge of compact low-field scanners is attracting the attention of an increasing number of laboratories worldwide. Cost effectiveness, portability, the lack of artifacts typically present at high fields, plus the possibility to perform secure diagnosis of certain events, even with a much lower image quality, is stimulating new developments and insights in this direction [31-33]. This picture can also be extended to FFC-MRI systems operating at limiting conditions: a minimum image quality enables the possibility to explore a plethora of physical and chemical processes that may turn into new specific contrasts. With improved performance, the system also turns into a cost-effective tool for pre-clinical studies using small animals (work in progress). With this motivation in mind, the main magnetic field of our prototype was generated using a variable geometry notch-coil magnet [34] whose magnetic field homogeneity was intentionally degraded. This procedure allowed us to analyze the compromise between the homogeneity of our magnet for a magnetic resonance image under field-cycled conditions (and its resolution) constrained by power supply performance.

A weak magnetic field inhomogeneity introduces a spatial distortion of the image that can be corrected if the field inhomogeneity map is known. However, the problem may become complex at higher inhomogeneities. Different mitigating strategies, at both during data acquisition or post-processing, were proposed to eliminate image artifacts originated in magnetic field inhomogeneity. Examples are methods with specific acquisition strategies of line segments in the k-space [35,36], a method where the contribution to the signal arising from the background magnetic field inhomogeneities is incorporated in the traditional gradient echo sequence (Voxel Spread Function or VSF approach) [37], a method using cross-sampling and self-calibrated off-resonance corrections [38], and a method analyzing phase shifts between adjacent pixels in different directions [39], among others. Moreover, two dimensional images of a flat object were obtained at highly inhomogeneous magnetic fields using a single sided device using a two-dimensional pulse sequence for phase encoding in the spin echo method [40]. So, we may conclude that the design of MRI devices operating at poor homogeneity conditions is highly feasible, while the effectiveness of the mentioned methods can also be analyzed in the context of the magnetic field-cycling approach.

It was already observed that different tissues associated with different anatomies in small animals (liver, kidney, spleen, fat, muscle, etc.) present different $T_1$ relaxation dispersions [41]. Experiments have been done by inserting a field-cycled electromagnet into a fixed-field main magnet, for example, in the magnet of an MRI scanner. The insert magnet may be bipolar to further increase the effective field range, a solution known as Delta Relaxation Enhanced Magnetic Resonance or DreMR [42-44]. A more recent patent proposed extending the method to other NMR parameters, not only limited to $T_1$ [45]. Other groups also explored similar approaches,as can be found in the literature [27,46,47]. A common feature is that a combination of a fixed magnet plus a field-shifting coil is used. While this approach offers a clear advantage in the attainable SNR ratio (the fixed magnetic field is usually a standard MRI magnet operating at 1.5 T or 3 T), plus the possibility of using the whole powerfulness of a conventional commercial MRI scanner (pre-programmed RF pulse sequence library, MRI-optimized hardware including the gradient unit, optimized magnet, optimized RF chain hardware, etc.), in general it also represents a high-cost solution for concept testing in contrast agent design, preclinical studies in animals, or applications in materials science.

A huge difference can be found in the proton spin-lattice relaxation dispersion slope between liver and fat, particularly at Larmor frequencies that are lower than 5 MHz. This tendency can also be observed between fat and other tissues like kidney, muscle and spleen [41]. On the other side, for the vast majority of the studied samples involving soft matter and porous materials, differences between free and constrained fluids, or corresponding to different phases, also occurs at proton Larmor frequencies that are usually lower than 5 MHz. It is then in this context that a fast-switchable moderate-cost instrument using a unique magnet, operating from a few Hz up to 5 MHz, capable to produce images of acceptable quality, may turn very attractive for testing concepts in designing physical or low-field specific chemical contrasts agents, for both biomedical applications or materials science.

## 2. The FFC-MRI Relaxometer

### 2.1 Block diagram

The system operates at maximum polarization and detection fields of 125 mT (5 MHz in proton Larmor frequency units). The spin-system can be left to evolve (or be manipulated) at a minimum Larmor frequency below 10 kHz. A typical slew-rate for the magnetic field between 11.6 T/s and 35 T/s was used for a maximum sample volume of 35 mL. The prototype was built with parts of our own design and fabrication, integrated into a modified Spinmaster relaxometer from Stelar (Mede – Italy). Particularly, the main electromagnet, the gradient unit, the RF probe and several peripheral electronic systems were specially designed and built at our laboratory. The main power supply, its cooling system and an auxiliary console are taken from the Stelar Spinmaster unit, conveniently modified to perform to our needs. The gradient amplifiers are Techron 8607 (Techron Division of Crown International Inc., Elkhart - USA), also modified to match our gradient-coil system. The main console is a Radio Processor-G board from Spin Core Technologies Inc. (Gainesville - USA). Fig. 1 presents a schematic block-diagram of the prototype.

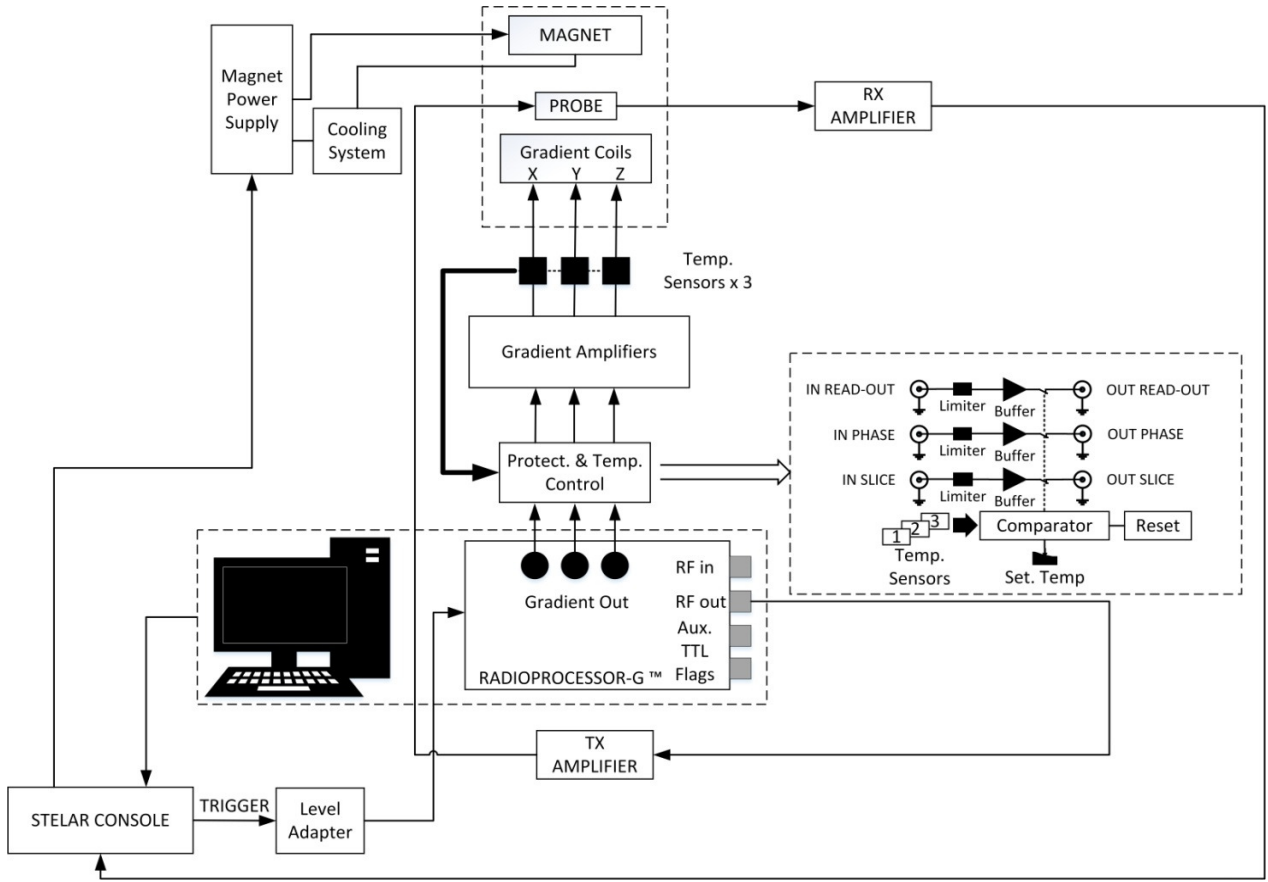

**Figure 1:** Complete schematic diagram of the prototype. Se text for details.

The transmitter included in the Stelar console was properly adjusted to amplify sinc-shaped pulses with a minimum distortion. The Techron 8607 gradient amplifiers were adapted to fit with the electrical characteristics of our gradient coils. For all gradient coils, the effective slew rate is above 1000 mT/m/ms, while switching times are less than 50μs (for the typical pulse amplitudes used in this work). A protection system was built to safeguard the gradient coils from excessive currents in case of malfunction. The auxiliary Stelar console controls the timing of the main magnetic field cycle and triggers the RF acquisition sequence that is controlled by the Radio Processor-G board. Extra peripheral hardware had to be fabricated for an adequate communication between the Stelar console and the Radio Processor-G board, and to match the impedances between the latter and the gradient amplifiers.

The main magnet is a variable geometry notch-coil composed of one layer and two external correcting coil-elements [34,48]. It produces a magnetic field of 125mT for a current of 170A, allowing switching from zero to the maximum field in less than 3ms without energy storage assistance. This electromagnet was designed under the premises of achieving an adjustable system through a variable geometry, capable of compensating homogeneity losses from, for example, mechanical degradation due to repetitive thermal stress.

The home-made gradient coil-system includes a longitudinal coil of our own design [49]. The transversal gradient coils were adapted from models already presented in the literature (see section S1 in Supplementary Material for details) [50,51]. The electrical parameters of the coils are shown in Table 1.

|   | $L$ [µH]   | $R$ [mΩ]    | $\eta$ [mT/m.A] |
|---|------------|-------------|-----------------|
| X | 68.5±0.2   | 367.3±0.1   | 9.0±0.1         |
| Y | 69.0±0.2   | 382.5±0.1   | 8.3±0.1         |
| Z | 98.7±0.2   | 572.5±0.1   | 15.9±0.1        |

**Table 1:** Measured parameters of the constructed magnetic field gradient coils. Values shown are: inductance (assembled gradient-system and placed inside the main electromagnet), resistance and coil efficiency.

**2.2 Magnet homogeneity and current stability**

For a cylindrical sample volume of 35 mL, the optimal magnet homogeneity is 130 ppm [34]. In this work however, the magnetic field homogeneity was set to 1400 ppm, since we were interested in the study of hardware simplifications that may be compatible with a reasonable quality of images. We show that a resolution better than 0.7 mm is possible in the images under such conditions. Degrading the magnetic field homogeneity has direct consequences in the requirements of the current stability, a point that in turn has a decisive impact in the complexity and cost of the power supply and the involved current control system. The 1400 ppm used in this work exceeds the maximum degradation that can be obtained by modifying the position of the notch coils (600 ppm). This value was obtained after a destructive testing of the magnet, where the homogeneity resulted originally in 4500 ppm, and after an optimization of the notch coil positions, in 1400 ppm. Details and aims of this test are outside the scope of this manuscript and will be presented elsewhere.

After the "passive shimming" of the magnet by optimizing the notch-coil positions (z-axis), a further improvement was performed by aligning the readout gradient along the maximum $B_0$ gradient in the plane. In this way, the linear component of the $B_0$ inhomogeneity gradient contributes to the readout gradient, and only the non-linear components of the inhomogeneity are effective. With this additional "active shimming", the effective inhomogeneity at signal acquisition was of 760 ppm.

The current stability of the power supply was determined from the standard deviation of the frequency shifts of the peak value of the Fourier transform (FT) of the echo signal (acquired without gradients, 50 samples). This measurement resulted in a value of 220 ppm. Since acquisition is performed at a Larmor frequency of 5MHz, the effective magnet homogeneity of 760 ppm governs the linewidth of the FT predicted as 3.8 kHz, as was verified experimentally using a sample of 35 ml of a water solution of copper sulfate 18 mM, $T_1$=(35±2) ms, $T_2$=(32±2) ms, both relaxation times measured at 5 MHz using an NP sequence. In the case of $T_2$ the NP sequence was combined with a Hahn echo experiment. Considering a central frequency of 5 MHz, the proportionality between current and generated magnetic field, and the 220 ppm of current stability, it turns out that the effective magnetic field stability is 1.1 kHz. A linewidth of 3.8 kHz provides a partial immunity to field fluctuations spanning over a 1.1 kHz bandwidth, allowing signal averaging with minimal signal loss.

In order to relate magnetic field fluctuations with the acquired echo-signal, we may consider the maximal instability of the field during the acquisition in the presence of the readout gradient [19]. If we want the accumulated phase of the magnetization during the acquisition time (during which the $G_{FE}$ gradient is switched on, see Fig. 2) to be small, we may consider the condition $\gamma \Delta B_0 T_{read} < \pi$. Here $\Delta B_0$ represents the standard deviation of the fluctuating field and $T_{read}$ the readout time during which the echo signal is recorded. The Nyquist criterion relates the spatial resolution with the readout parameters so that the previous condition can be written as $\Delta B_0 < G_{FE} \Delta x$, that is, field fluctuations should be smaller than the pixel bandwidth. In our case we may consider a typical readout gradient of 30 mT/m and a pixel size $\Delta x = 1$ mm. This implies a required field stability of 240 ppm that is in fact worse than the measured stability of the current (220 ppm). The fact that we do not observe undesirable ghosting in the images is also consistent with the fact that the effective field inhomogeneity is larger than the field fluctuations bandwidth.

## 2.3 Pulse sequences and imaging parameters

The spin-echo sequence was chosen for image acquisition due to its robustness under the presence of $B_0$ inhomogeneities [52]. This sequence is combined with either pre-polarized (PP) or non-polarized (NP) field-cycles (Fig.2). The region colored in light gray in Fig. 2 (before the 90-degree RF pulse) is used to stabilize the magnetic field gradient and also to dephase any remaining transverse magnetization (spoiler gradient). Appropriate gradient parameters must be chosen to effectively rephase all spins across the selected slice; otherwise signal intensity can be severely degraded [53].

Magnetic field inhomogeneities of the magnet introduce spatial encoding that, if not adapted for in the image reconstruction process, produce distortion and artifacts. Thus, a correction of such inhomogeneities is needed to preserve the image quality. The solution used in this work consisted in lining-up the readout gradient along the main component of the inhomogeneity in the transaxial plane, while setting a high gradient intensity in this direction [54].

The sinc-shaped RF pulses used here had three lobes and were apodized with a Hann function. Their duration was 400 μs and the resulting irradiation bandwidth was 20 kHz. The echo time (*TE*), gradient pulse durations and their intensities were changed between experiments.

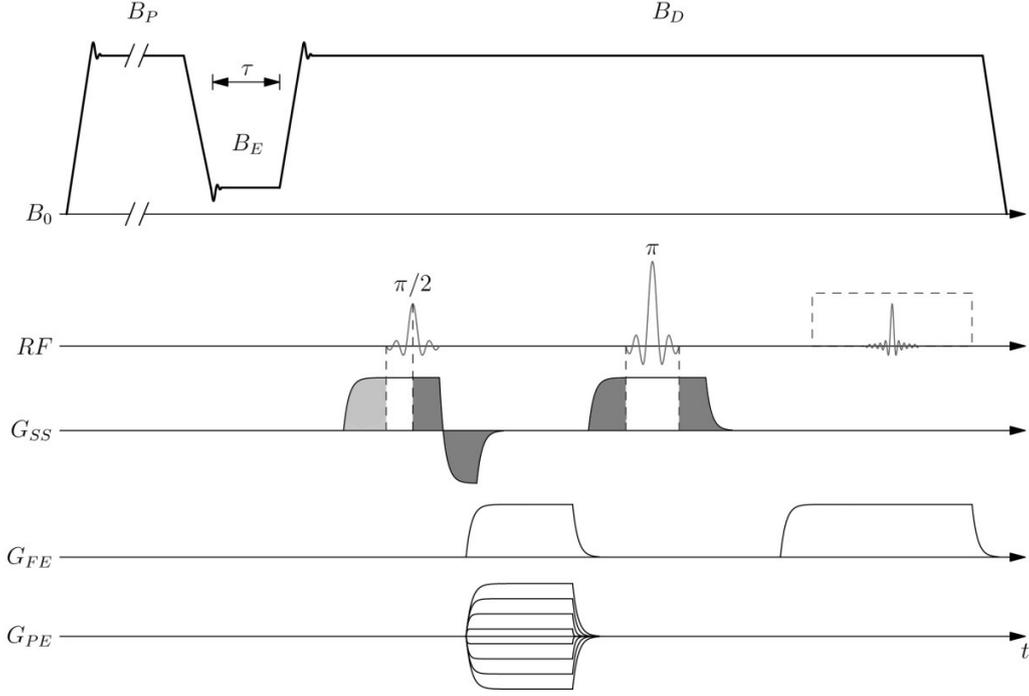

**Figure 2:** Pre-polarized (PP) FFC-MRI sequences used in this work. RF: Radio Frecuency. SS: Slice Selection. FE: Frecuency Encoding. PE: Phase Encoding. $B_P$: polarizing field. $B_E$: evolution field. $B_D$: detection field. $\tau$: evolution time. In the SS sequence, regions with same color must have equal area. The NP variant of the sequence only differs in that no polarization field is applied at the beginning (i.e., $B_p=0$).

## 3  Experiments: testing the FFC-MRI relaxometer capabilities

The accomplished experiments were chosen to demonstrate the basic capabilities of the instrument operating at the given conditions. Measurements of $T_2$ (at a Larmor frequency of 5 MHz) corresponding to the different samples were performed using the same instrument. A field pulse of 2s length and a Hahn echo sequence was used in all cases. The SNR in all images was calculated as the ratio of the mean value of the intensity within the ROI (region of interest) and the standard deviation of the rest of the image not including the ROI.

### 3.1 Image resolution at maximum SNR (without slice selection)

Fig. 3 shows an image obtained using a NP sequence without slice selection, i.e., a transaxial projection alongside the whole sample length (30 mm). In this way we can reach the maximum SNR ratio to determine the image resolution, while testing the sample alignment with the gradient unit. To further improve the image quality, a post processing filter was implemented. The *unbiased* non-local means (UNLM) algorithm was chosen because it has been proven to outperform other commonly used denoising methods [55]. The shading across the phantom is originated in phase noise ghosting that becomes spread over the image by the digital filtering. The resolution is of the order of 0.8 mm (which is the thickness of the glass tubes). Pixel spreading is position dependent due to the image distortion originated in the magnetic field inhomogeneity. We can observe that a physical dimension of 0.8 mm x 0.8 mm corresponding to the phantom (at the glass tubes walls for example) spreads out into an average effective pixel of 1 mm x 1 mm. This fact comes as an encouraging result for an image obtained at a magnetic field of 1400 ppm homogeneity and a current stability of 220 ppm.

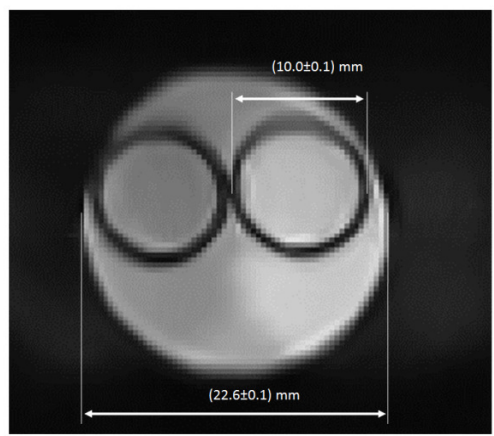

**Figure 3:** 2D image acquired without slice selection. The phantom consists of a cylindrical volume of water doped with copper sulfate 18mM, $T_1$=(35±2) ms, $T_2$=(32±2) ms (both relaxation times measured at 5 MHz) containing 2 glass tubes of 0.8 mm wall-thickness

immersed in it. Image parameters: matrix size 128x128, 4 scans, $G_{read} = 84.0\ mT/m$, $G_{phase}^{MAX} = 49.3\ mT/m$, $TE = 2\ ms$, FOVy= 45 mm, FOVx = 44 mm, total acquisition time 18 min. Repetition time (TR) is 2s and $\tau$ = 120 ms. SNR=850.

### 3.2 Volume-selective $T_1$ measurements in heterogeneous samples

A feature arising from the merge of the FFC and MRI techniques is the possibility to measure $T_1$ dispersions with spatial localization in small volumes within a sample. Without the aid of imaging techniques, such experiments can only be performed using surface coils. This approach may require physical access to the interior of the sample, which can be a complicated procedure while resulting in sample contamination. When working with samples "in vivo", this may imply a surgical incision [56]. In contrast, the use of magnetic field gradients for spatial encoding and volume selection is a direct and non-invasive practice. Similar techniques are used in localized NMR spectroscopy [57,58]. This approach has already been used, employing a permanent magnet equipped with an electromagnet for cycling the magnetic field (baseline relaxometry) [29]. The method to determine the minimum selected volume and its uncertainty is discussed in section S2 of the Supplementary Material.

A sample holder consisting of two isolated chambers was built to test the capabilities of our machine for 1D spatial localization (slice-selection). Each chamber was filled with deionized water doped with copper sulfate at different concentrations (1.1 and 11 mM), resulting in a binary sample with 2 different spin-lattice relaxation times. The copper sulfate was from Cicarelli (San Lorenzo, Santa Fé, Argentina) and the deionized water obtained from a milli-Q equipment Osmoion 5 from Apema (Villa Dominico, Buenos Aires, Argentina). A non-polarized (NP) field-cycled pulse sequence consisting of a spin-echo with sinc-shaped RF pulses synchronized with slice-selection field gradients was used. The sequence timing was simmilar to the scheme shown in Fig. 2, but with $B_P=0$. All experiments were done at 22 ºC.

The longitudinal gradient (z) was used to select transaxial planes of each subsample individualy, whilethe transversal gradient (x) was used to select a sagital plane emcompassing both subsamples. The corresponding volumes of sample within the selected slices are 1.5 mL for both transaxial slices (4.6 mm thickness) and 2 mL for the sagital slice (4.8 mm thickness). The experiment allows checking the capability of the system in identifying each $T_1$ component in the sagital selection, i.e., in the context of a non-homogeneous sample (in this manuscript we define a "$T_1$ component" as a partial volume of the sample relaxing with a given $T_1$). Fig.4 shows the corresponding $T_1$ measurements with an ilustration of the selected slices. Each measurement was repeated 16 times to assess reproducibility. The fit of the biexponential curve was obtained using the measured $T_1$ values of the individual subsample volumes (regions), giving a $R^2$ of 0.9967.

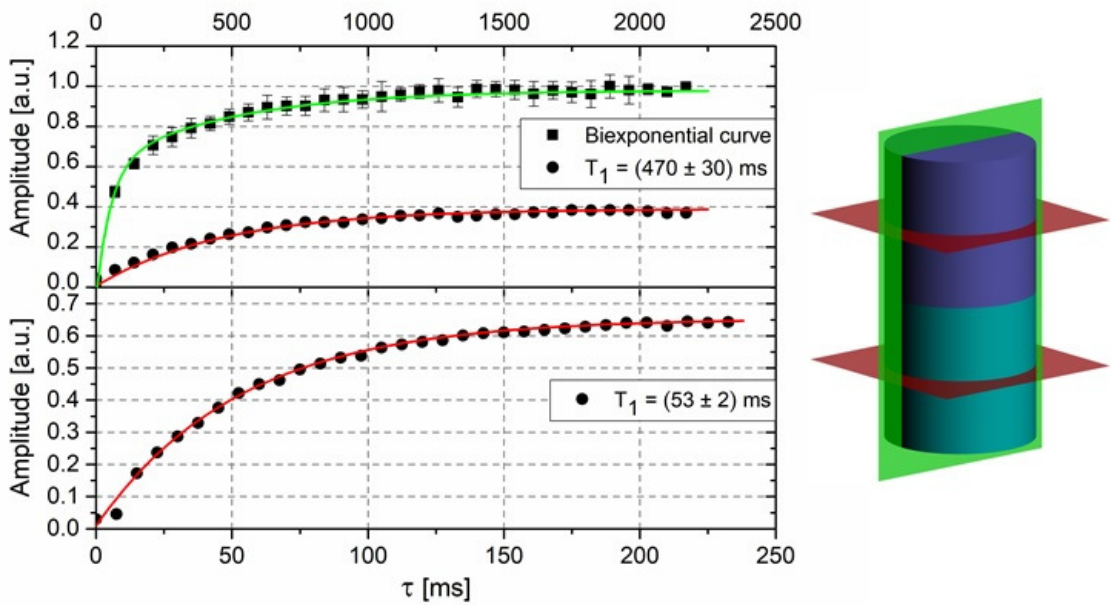

**Figure 4:** Magnetization evolutions corresponding to the different slices: **A** (sagittal) and **B** and **C** (transaxial, each located within a different subsample). The magnetization evolution corresponding to **A** (squares) show a biexponential behavior (see text for details). $T_1$ measurement of B (deionized water doped with 1.1 mM of copper sulfate) resulted in (470 ± 30) ms, within 95% confidence interval. $T_1$ measurement of **C** (deionized water doped with 11 mM of copper sulfate) gives (53 ± 2) ms, also within 95% confidence interval. Magnetization evolutions corresponding to the transaxial planes are indicated with circles, and the corresponding fitting in red. *TE* = 2 ms and *TR* = 2.5 s. 16 scans (following a dummy scan) were accumulated to determine each magnetization value (with a receiver bandwidth of 30 kHz).

### 3.3 $T_1$ dispersion-weighted images

The magnetization associated with a given voxel at the end of the relaxation period depends on the time $\tau$ during which the spin system evolve at the relaxation magnetic field $B_R$ and, the longitudinal relaxation time $T_1(B_R)$. The relaxation field defines the Larmor frequency $\nu_0$ at which the spin-lattice time is measured, and the function $T_1(\nu_0)$ represents the $T_1$ dispersion curve [13]. In heterogeneous samples different voxels may be associated with different spin-densities, but they can also relax according to different $T_1$ dispersions. If we consider a case with two relaxation components showing different $T_1$ dispersions, the contrast can be maximized by acquiring the images at the Larmor frequencies where differences in the involved relaxation times are maximal.

Understanding of the capabilities of the instrument has been gained for acquiring images using $T_1$ dispersion-weighted contrast. Test experiments were implemented using PP and NP sequences in a phantom consisting of two tubes containing Polydimethylsiloxane (PDMS, molecular weight $M_w$=5200, melt phase, PSS Mainz - Germany) as a dispersive component, and two tubes containing an aqueous solution (AS) of CuSO4 at 2.5 mM concentration (non-dispersive component). Dispersion curves for each individual component were measured with a Stelar Spinmaster FC2000/C/D relaxometer. The contrast obtained using the prototype field-cycling imager was compared with the contrast derived from the measured dispersions through a simple model (details in section S3 of the Supplementary Material).

Fig. 5 shows a 2D image of the phantom without slice selection (sample length is 30 mm). The concentration of the aqueous solution of CuSO$_4$ was adjusted in order to have the same $T_1$ as the PDMS melt at 5 MHz: $T_1$ = (240±10) ms. The measured value of $T_1$ corresponding to PDMS at 10 kHz was $T_1$ = (20 ± 1) ms, while for the aqueous solution at the same Larmor frequency was $T_1$ = (220 ± 10) ms. For PDMS $T_2$ = (1.9 ± 0.1) ms and for the aqueous solution $T_2$ = (73 ± 9) ms, both measured at a Larmor frequency of 5 MHz. Since at 5 MHz both samples exhibit approximately the same spin lattice relaxation time, the contrast obtained at 5 MHz using a NP sequence reflects the difference in proton density and $T_2$ (Fig. 5A). A maximum contrast can be obtained by using a PP sequence with parameters set in such a way that the magnetization associated with the dispersive component can be made negligible, while the magnetization from the aqueous sample has only slightly decreased (Fig. 5B). ). In this case the aqueous sample produces a positive PP contrast.

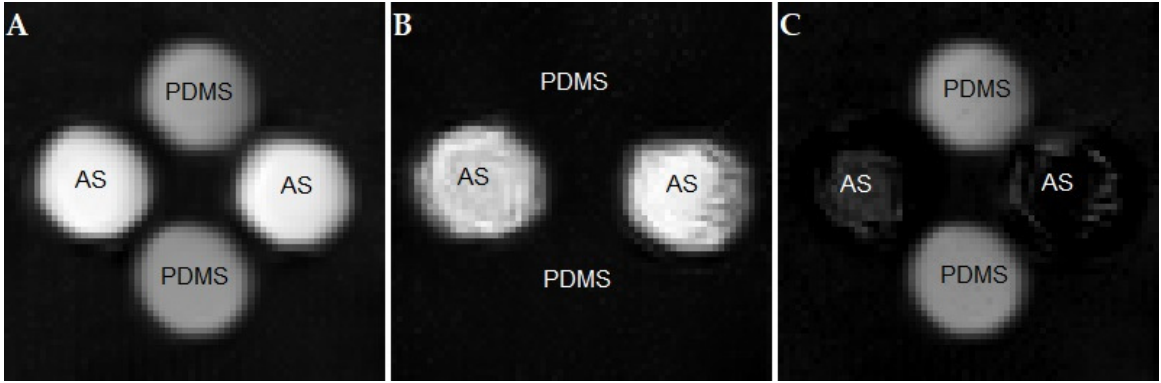

**Figure 5:** Magnitude images of a phantom consisting of two tubes containing a PDMS polymer melt and two tubes filled with an acqueous solution (AS) at 22 °C, see text for details. (A) Image acquired using a NP sequence with $\tau$= 960 ms at 5 MHz. The bright difference berween PDMS and AS tubes correspond to a difference in the spin density and $T_2$. SNR=850. (B) PP positive contrast: image acquired using a PP sequence with $B_E$ = 10 kHz and $\tau$= 48 ms. The PDMS component has completely disapeared. In contrast, the signal intensity loss due to relaxation for the acqueous component is less than 15% (average). SNR=290. (C) $T_1$ dispersion-weighted contrast: Difference between images A and B. SNR=111. In all cases image parameters are: matrix size 64x64, 2 scans, $G_{read} = 42\ mT/m$, $G_{phase}^{MAX} = 29\ mT/m$, $TE = 2\ ms$, FOVy= 28 mm, FOVx= 25 mm, total acquisition time 4.7 min. $TR = 1.2\ s$.

It is also possible to obtain the inverted contrast, where the dispersive component is preserved and the aqueous component filtered, by taking the difference between the NP and PP acquired images (Fig. 5C). In this case, the dispersive component is associated to a positive contrast ($T_1$ dispersion-weighted contrast).

### 3.4 Inverted $T_1$-weighted images

It may happen that in a given sample having two or more components none of them is dispersive. This would be the case of samples containing free water in excess in different compartments, each of them probably having a different $T_1$.

When at least two $T_1$ components are clearly distinguishable, the NP sequence can be used to maximize the *inverted* $T_1$ contrast in a short time. In this case, the magnetization evolution of each $T_1$ component during the relaxation period can be expressed as:

$$M(t) = M_0(1 - e^{-t/T_1}). \tag{1}$$

Let us assume for simplicity that only two distinguishable components are considered. Writing the two spin-lattice relaxation times as $T_1^{(1)}$ and $T_1^{(2)} = kT_1^{(1)}$ with $k>1$, after a little algebraic effort it can be shown that there exists an optimum relaxation delay $\tau^*$ where the contrast is maximum:

$$\tau^* = \frac{kT_1^{(1)}\ln(k)}{k-1}. \tag{2}$$

An example is discussed in detail in the Supplementary Material (Section S4).

**3.5 Images in a fat/muscle interface of bovine tissue**

Dispersion-weighted contrast imaging was tested in a sample of bovine tissue (sample extracted from between the upper front ribs and the skin obtained from a butcher shop) at an interface of fat and muscle. Measured values of $T_2$ at 5 MHz and 22 ºC were $T_2$ = (42 ± 4) ms for muscle and $T_2$= (24±2) ms for fat. Dispersion curves of fat and muscle tissue were measured independently at 22 ºC to gather information on their relaxation times (see Fig.6). Muscle fiber has a more dispersive behavior than fat and their curves cross at approximately 100 kHz. It is important to notice that relaxation times for both samples increase monotonically with the effective Larmor frequency that pertains during the evolution period. Difference between curves also increases with the operating Larmor frequency, which should improve contrast using a conventional scanner with the inversion recovery technique. On the other hand, contrast also improves for lower fields, where relaxation times are much shorter.

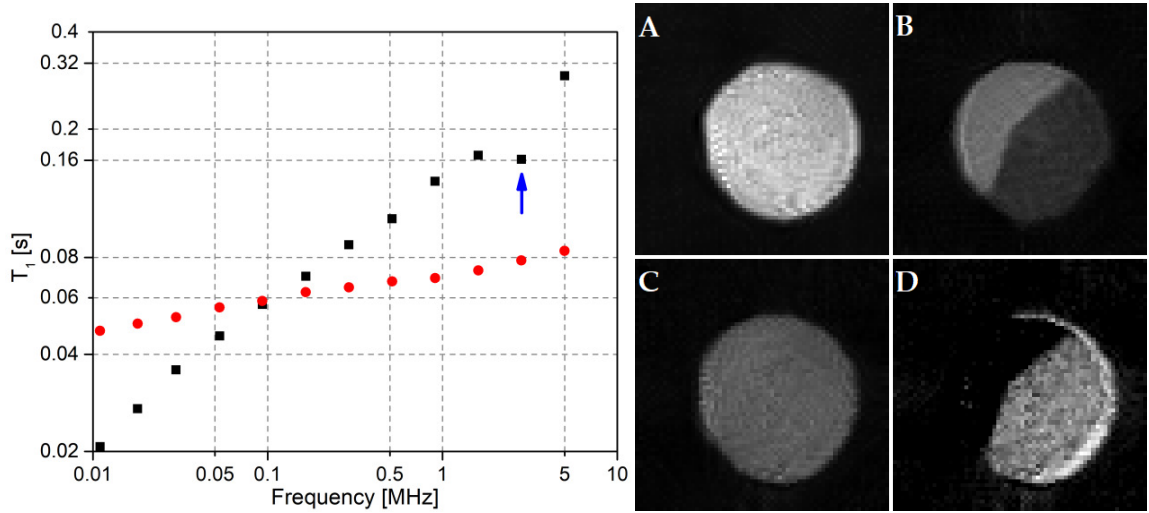

**Figure 6:** At left, dispersion curves corresponding to muscle (squares) and fat (circles) of bovine tissue. The local minimum in the muscle tissue curve at approximately 2.8 MHz corresponds to a quadrupolar dip (indicated with an arrow). The difference between both relaxation dispersion curves increases at Larmor frequencies higher and lower than about 100 kHz (where a crossover of the two dispersion curves occurs). At right, 2D images of a transaxial slice selection of the bovine tissue sample at a muscle/fat interface. All images were acquired using 64x64 data matrices, slice thickness of 4 mm and 8 averaged signal scans. (A) Image with NP sequence, $B_E$ = 5 MHz, $\tau$ long enough (1.5s) to saturate the magnetization of both components. SNR=267. (B) Image with inverted $T_1$-weighted contrast. SNR=96. (C) Image acquired with a PP sequence, $B_E$ = 100 kHz and $\tau$ = 30 ms. SNR=120. (D) Renormalized image difference obtained from images (C) and (B). SNR=18. In Fig. (D) only the muscle tissue is observed. $G_{read} = 31.5 \; mT/m$, $G_{phase}^{MAX} = 32.7 \; mT/m$, $TE = 2 \; ms$, FOVy= 36 mm, FOVx= 38 mm, $G_{SS} = 126 \; mT/m$, $TR = 1.4 \; s$.

A first image was acquired using a NP sequence at 5 MHz with a long $\tau$ (1.5s) in order to saturate the magnetizations of both components (Fig. 6A). The resulting image exhibits a uniform intensity, which it is not possible to distinguish between muscle and fat (both components have approximately the same proton density). Contrast however can be generated using the information contained in the relaxation dispersion curves shown at the left side of Fig.6. In Fig. 6B (inverted $T_1$-weighted) the component with the lower $T_1$ value (fat) turns bright in the image (NP positive contrast). It is also possible to obtain the inverse contrast: an image where the component with the higher $T_1$ value (muscle) looks bright. To do this a third image was acquired using a PP field-cycling sequence with $B_E$= 100 kHz. It is necessary to acquire the image with the exact $\tau$ value such that the fat signal intensity has the same value as that in Fig.6B. For this purpose, the optimal $\tau$ value used for the image of Fig.6B was employed, and an exponential decay of magnetization was assumed. The optimal $\tau$ value was then calculated by forcing the magnetization intensities from the fat component to be the same in both cases. Fig. 6C shows the acquired image and Fig.6D the normalized difference between images 6B and 6C. In this case the contrast is dominated by the component with higher dispersion (muscle), while the less dispersive component (fat) was filtered. All images were acquired at 22 °C.

## 3.6 Temperature Mapping

Temperature maps can be imaged using MRI [59-61]. The non-invasive and high-image-resolution features of the technique make it particularly suitable for clinical applications such as hyperthermia treatment [60-62]. There are a number of physical parameters that exhibit a dependency with temperature, but the most widely studied so far have been water proton chemical shift [63], diffusion coefficient [64] and $T_1$ relaxation time [59]. MRI temperature mapping using the dependency of $T_1$ with temperature has never been studied in the context of FFC.

For these experiments a 20mL sample of gelatin from porcine skin (Fluka, Buchs – Switzerland) diluted in Milli-Q water at 39% concentration was used. A series of $T_1$ dispersion curves at different temperatures were measured using a commercial relaxometer (Stelar Spinmaster FC2000/C/D). The series of measurements covered the range from 30 to 60 °C at 5 °C steps (Fig.7A).

A NP sequence with $B_E$ = 5 MHz was used to minimize acquisition time. An optimum relaxation delay $\tau^*$ of 735 ms was used to maximize the temperature-dependent contrast between 30 and 45 °C (see Section 3.4). This scheme produces temperature maps where the lowest signal intensity corresponds to the highest temperatures. The opposite is also possible (highest signal intensity for highest temperature), but requires a PP sequence and consequently, longer acquisition times. After image acquisition, temperature measurements can be obtained by using Fig.7B and inverting equation (2) to get:

$$T_1 = \frac{-\tau^*}{ln\left[1-\frac{M(\tau^*)}{M_0}\right]}. \qquad (3)$$

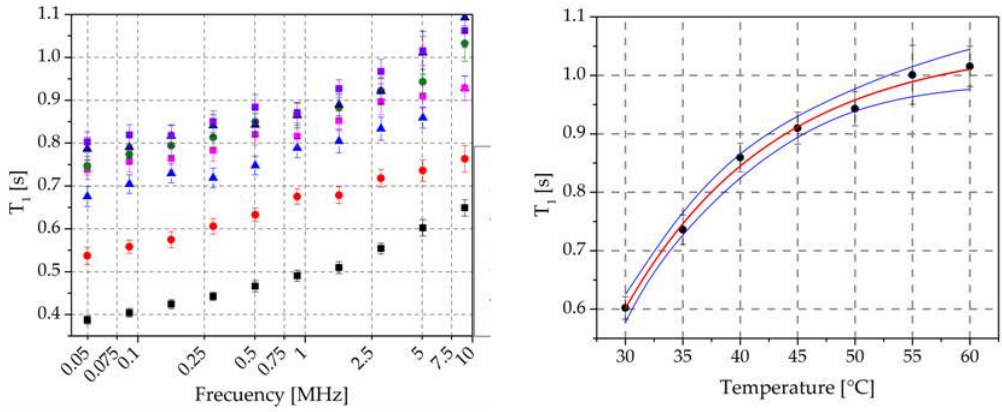

**Figure 7:** (A) Dispersion curves of porcine gelatin sample at different temperatures. Color legend to the right refers to temperature of the sample in °C. Error in temperature calibrations are within ±1 °C. (B) Temperature dependence of $T_1$ at 5 MHz. The fitted curve (red) was obtained using the function $T_1 = A[1 - \exp(k(\theta - \theta_C))]$ with $R^2 = 0.994$, where $\theta$ is the measured temperature, $A = (1.06\pm0.02)$ s, $k = (0.07\pm0.01)$ °C$^{-1}$ and $\theta_C = (19\pm1)$ °C. Upper and lower 95% confidence intervals are shown in blue.

A 20 mL sample of porcine gelatin was placed in a glass tube of 21mm inner wall diameter. The sample was heated from the center (coaxially with the azimuthal axis of the sample) using a home-made glass "cold finger". The device has an outer diameter of 7 mm and its temperature is controlled by circulating a flux of hot water through it. After the temperature was stabilized throughout the sample volume (approximately 15 minutes) a small persistent radial temperature gradient could be observed. For the highest temperature (45°C) the amplitude of the persistent temperature gradient was 3°C (temperature difference between the outer wall of the "cold finger" and the inner wall of the sample holder tube). These measurements were done with thermocouples using a CHY 503 electronic thermometer (CHY Firemate, Tainan - Taiwan). Four images were acquired at sample temperatures of 30, 35, 40 and 45 °C (Fig. 8A). Fig. 8B shows the comparison of temperature vs. $T_1$ plots obtained from the MRI experiment and from direct temperature measurements on the sample. The signal intensity from an image acquired at room temperature was used to determine $M_0$ (used in equation (3)).

In this experiment we have neglected the temperature dependence of $T_2$. Note that in general for gelatins $T_2$ increases with the temperature [65,66]. Provided that $T_2 = (31 \pm 3)$ ms at 30 ºC (5 MHz) and TE = 2 ms, the increase in $T_2$ at higher temperatures cannot be observed within experimental errors (see Fig. 8B).

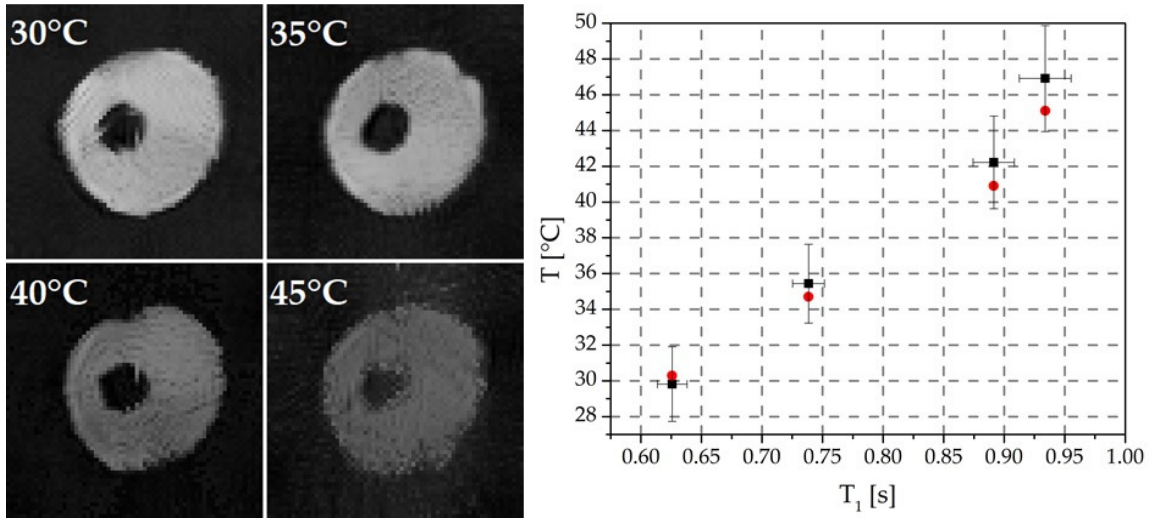

**Figure 8:** (A) Temperature map images of the porcine gelatin sample. The black circle at the center of each image is the "cold finger" used to heat the sample. No proton signal is observed within this region due to the fact that the heating water is flowing: polarized protons are outside of the FOV at the time the RF excitation is applied. All images correspond to a slice thickness of 4 mm. A NP sequence was used in each case, with $B_E$=5 MHz and $\tau^*$= 735 ms to maximize the contrast between $T_1$ values at sample temperatures 30 and 45 °C, according to equation (2). $T_2$ effects are neglected (see text for details). (B) Temperature vs. $T_1$ plot obtained from a) MRI method (black squares) and b) invasive thermocouple measurements (red circles) close to the inner wall of the sample holder (lowest point of the radial temperature gradient). Images parameters are: matrix size 64x64, 8 scans, $G_{read} = 31.5\ mT/m$, $G_{phase}^{MAX} = 32.7\ mT/m$, $TE = 2\ ms$, FOVy= 36 mm, FOVx= 38 mm, total acquisition time 16 min. $G_{SS} = 126\ mT/m$. SNR: 368 (30 ºC), 361 (35 ºC), 125 (40 ºC) and 95 (45 ºC).

## 4 Discussion and final remarks

Advantages of homogeneous fields are the favorable SNR and the access to spectroscopic resolution. When this option is implemented using electromagnets, high requirements are imposed to the power supply in order to minimize thermal shifts and short term instability of the current [67]. For MRI and spin-echo applications, a current stability much higher than the magnetic field inhomogeneity bandwidth is usually a prerequisite. Current (and field) instability causes phase noise that introduces artifacts and ghosting in the image, a problem that can be mitigated through specific post-processing filters [68]. However, hardware costs can be strongly reduced by relaxing homogeneity and stability requisites. Undesirable effects due to field instabilities and inhomogeneities can be minimized with an adequate pixel bandwidth, for instance, by increasing the intensity of the readout gradient. This has a negative impact in the SNR that can be compensated by accumulating a higher number of acquired echoes. In addition, many strategies at both during sampling and post-processing can be implemented to correct for image distortions and undesirable artifacts. Therefore, we may conclude that MRI at modest homogeneity and stability conditions can be manageable at the cost of longer acquisition times, being this also applicable for field-cycling MRI.

A main difference between the prototype here presented and other FFC-MRI machines described in the literature relies in the approach: here we present a relaxometer with MRI capabilities, conceived as a laboratory tool for the design of physical and chemical contrasts of specific use in FFC-MRI scanners. A new funding stage of this project is focused in improving the hardware and software to allow better-quality images, with potential use in pre-clinical studies in small animals. In this context, several instrumental and software developments are feasible to improve the SNR and image quality at low fields [38,69].

In this work we show the feasibility of obtaining FFC-MRI images under conditions of poor magnetic field homogeneity. This fact relaxes the power supply and control electronics requirements, especially on stability demands and concerning the magnet design as well. This opens the possibility for low-cost devices that can be adopted by laboratories working in pharmacology, preclinical studies, development of contrast agents and materials science, among others.

The machine here presented is based on a commercial relaxometer from Stelar Srl (Mede-Italy) that was conveniently modified for the mentioned purposes. A gradient unit was designed and fabricated according to the prototype requirements. Peripheral hardware was introduced to complete the system. The Non-Local Means filter was successfully implemented. Experiments were carried out to calibrate the system as a whole, and experiments were presented to test the machine capabilities under unfavorable conditions.

An effective magnetic field inhomogeneity of 760 ppm was obtained by conveniently aligning the readout gradient along the maximum inhomogeneity gradient in the transversal plane (normal to the magnet symmetry axis). In this way, only non-linear components of the inhomogeneity in this direction are effective during the echo signal acquisition. At the light of the present results, the used magnetic field homogeneity and stability are enough to produce images of reasonable quality for the design and concept testing of contrasts for FFC-MRI.

The overall performance of the prototype was validated by performing $T_1$ relaxometric measurements and acquiring 2D images with slice selection. The minimum sample volume for which $T_1$ measurements can be achieved within a 30% precision was found to be 0.25 mL (20% for 0.5 mL, more details in the Supplementary Material). Temperature map imaging was tested for the first time in the context of field-cycling MRI.

Different contrasts were discussed and shown experimentally. They were both defined in terms of $T_1$ (in opposition to standard MRI where in general a positive contrast is categorized positive for short $T_1$ and negative for short $T_2$). In general, a positive contrast is associated to a higher intensity in the image. In a PP sequence this corresponds to longer $T_1$ components (PP positive contrast) while the contrary holds for a NP sequence (NP positive), i.e. associated to short $T_1$ components. A negative contrast is then associated with low-intensity in the image, that is, short $T_1$ components in a PP sequence (PP negative contrast), and with long $T_1$ components in the NP case (NP negative contrast). Note that for $T_1$ dispersion-weighted contrast, in the image resulting from subtraction of images acquired with a PP and a NP sequences, the positive contrast is associated to field-dependent (dispersive) $T_1$ components. In this case, a positive contrast is similar to that defined for a DREMR experiment [42].

Temperature maps were acquired at 5 MHz using a NP sequence. This experiment shows that the machine can also be used at "fixed-field mode" by switching the magnetic field on during all the needed time to polarize the spin-system and collect the data. Temperatures determined by inserting a thermocouple in the phantom and calculated from the NMR experiment are consistent. A PP sequence or a $T_1$-dispersion weighted contrast can be used for samples where dispersive components are temperature dependent.

Further improvements on this prototype are directed towards improving image quality and resolution at limiting homogeneity and current stability conditions, using just one switched magnet. The instrument is already useful to explore innovative physical contrast agents based on low-field (even close to zero) manipulation of the spin system. Details of these experiments and further improvements on the hardware and image processing will be shown elsewhere.

**Acknowledgements**

The authors acknowledge funding from FONCYT PICT-2013-1380 and 2017-2195, Province of Córdoba (PID and PROTRI) and Secretaría de Ciencia y Tecnología, Universidad Nacional de Córdoba (Secyt-UNC) for supporting this project. Fellowships and positions from CONICET are also acknowledged. The authors gratefully acknowledge help from Eng. Alexis Berté for collaborating in the design and implementation of different auxiliary hardware used in this work. We also acknowledge Dr. Gabriela A. Dominguez for helping in the set-up of the gradient system.